\documentstyle[12pt]{article}
\begin{document}

\title{Exponential Type Complex and non-Hermitian Potentials in $PT-$ Symmetric
Quantum Mechanics}
\author{\"{O}zlem\ Ye\d{s}ilta\d{s} $^{a}$, \ Mehmet \d{S}im\d{s}ek
$^{a}$, Ramazan Sever$^{b}$, Cevdet Tezcan $^{c}$ \\
{\small {\sl $^a$ Department of Physics, Faculty of Arts and
Sciences, Gazi University ,}} {\small {\sl 06500, Ankara ,
Turkey}}\\ {\small {\sl $^b$ Department of Physics, Middle East
Tecnical University ,}} {\small {\sl 06531, Ankara , Turkey}}\\
{\small {\sl $^c$ Department of Physics, Faculty of Sciences,
Ankara University,}} {\small {\sl 06500, Ankara , Turkey}}}
\date{}
\maketitle

\begin{abstract}
Using the NU method [{\it A.F.Nikiforov, V.B.Uvarov, Special Functions of
Mathematical Physics, Birkhauser,Basel,1988}], we investigated the real
eigenvalues of the complex and/or $PT$- symmetric, non-Hermitian and the
exponential type systems, such as P\"{o}schl-Teller and Morse potentials.
\end{abstract}

\baselineskip=22pt plus 1pt minus 1pt

\vspace{0.5cm}



\bigskip \noindent Corresponding author: R. Sever, \noindent
\noindent E-mail: sever@metu.edu.tr
\newpage

\section{Introduction}

Recently, an important feature of the $PT$-symmetric Hamiltonians is
recognized. So that they have real spectra although they may be Hermitian or
not. Initial studies of Bender and his co-workers on the $PT$ - symmetric
quantum mechanics which was restricted for those eigenvalues of Hermitian
operators which are real. Physical motivation for the $PT$-symmetric but
non-Hermitian Hamiltonians have been emphasized by many authors \cite
{Bender,Bender2,cannata,khare,bag,ZA,BQ,S,AM,Znojil,BA}. Following these
detailed works, non-Hermitian Hamiltonians with real or complex spectra have
been studied by using numerical and analytical techniques~\cite
{cannata,khare,bag,ZA,BA,bb}. The interesting results of non-Hermitian
quantum mechanical $~PT-$symmetric theories have been extended to the
two-particle bound states of conventional $g\phi ^{3}$ and -$~g\phi ^{4}$
quantum field theories\cite{BS}, and also to the two-point Green's function
representations\cite{bbm}.

There exist some physical reasons for such a generalization based on the
time reversal operator $T$ which is antilinear in the complex theory. Thus,
the operator $PT$ commutes with the Hamiltonian H. There are well known
physical parameters\cite{JM} saying that the Hilbert space , ~$%
L_{2}(x_{1},x_{2})$, of quantum mechanics could be real or complex in the
interval $~x_{1}<x<x_{2}\cite{Dyson62}~$. In quantum mechanics,
complexification of the energy can be used to describe resonance cases and
scattering theory \cite{Newton}.

In this work, the Schr\"{o}dinger equation is solved by using Nikiforov -
Uvarov (NU) method\cite{NU} to get energy eigenvalues of bound states for
real and complex form of \ the potentials. For the numerical application,
general forms of Morse and P\"{o}schl-Teller potentials are solved. One can
also use Cauchy-Riemann equations in the complex space, with the additional
degrees of freedom to study same complex Hamiltonian systems\cite{KK}.

After a brief introductory discussion of Nikiforov - Uvarov (NU) method\cite
{NU} in Section 2, we study complex and/or $PT$- symmetric non-Hermitian
exponential type systems with the P\"{o}schl-Teller and Morse potentials in
Section 3 and 4. Results are discussed in Section 5.

\section{Nikiforov-Uvarov Method}

NU method\cite{NU} is based on the solutions of general second order linear
equation with special functions. In this method, the Schr\"{o}dinger
equation in one dimension can be written in general as
\begin{equation}
\psi ^{\prime \prime }(s)+\frac{\stackrel{\sim }{\tau }(s)}{\sigma }(s)\psi
^{\prime }+\frac{\stackrel{\sim }{\sigma }(s)}{\sigma ^{2}(s)}\psi (s)=0
\end{equation}
where$~s=s(x)~$, $~\sigma (s)$, and $~\stackrel{\sim }{\sigma }(s)~$ are
polynomials, at most second - degree, and $~\stackrel{\sim }{\tau }(s)$~is a
first - degree polynomial. In the NU method the new function~$\pi ~$ and the
parameter~$\lambda ~$ are defined as
\begin{equation}
\pi =\frac{\sigma ^{\prime }-\stackrel{\sim }{\tau }}{2}\pm \sqrt{(\frac{%
\sigma ^{\prime }-\stackrel{\sim }{\tau }}{2})^{2}-\stackrel{\sim }{\sigma }%
+k{\sigma }}
\end{equation}
and
\begin{equation}
\lambda =k+\pi ^{\prime }.
\end{equation}
On the other hand, in order to find the value of $k$, the expression in the
square root must be square of polynomial. Thus, a new eigenvalue equation
for the Schr\"{o}dinger equation becomes
\begin{equation}
\lambda =\lambda _{n}=-n\tau ^{\prime }-\frac{n(n-1)}{2}\sigma ^{\prime
\prime },~~~(n=0,1,2,...)
\end{equation}
where
\begin{equation}
\tau (s)=\stackrel{\sim }{\tau }(s)+2\pi (s),
\end{equation}
and it will have a negative derivative\cite{NU}. The wave function is
constructed as a multiple of two independent parts,
\begin{equation}
\psi (s)=\phi (s)y(s),
\end{equation}
where ~$y(s)~$ is the hypergeometric type function which is described with a
weight function $\rho $ as
\begin{equation}
y_{n}(s)=\frac{B_{n}}{\rho (s)}\frac{d^{n}}{ds}\left[ \sigma ^{n}(s)\rho (s)%
\right] ,
\end{equation}

where $\rho (s)$ must satisfy the condition \cite{NU}
\begin{equation}
(\sigma \rho )^{\prime }=\tau \rho .
\end{equation}
Other part is defined as a logarithmic derivative
\begin{equation}
\frac{\phi (s)^{\prime }}{\phi (s)}=\frac{\pi (s)}{\sigma (s)}.
\end{equation}

\section{Generalized Morse Potential}

We shall first study different types of the general Morse potential,
\begin{equation}
V(x)=V_{1}e^{-2\alpha x}-V_{2}e^{-\alpha x}.
\end{equation}
Now, in order to apply the NU-method, we write the Schr\"{o}dinger equation
with the generalized Morse potential by using a new variable ~$s=\sqrt{V_{1}}%
e^{-\alpha x}$%
\begin{equation}
\psi ^{\prime \prime }+\frac{1}{s}\psi ^{\prime }-\frac{1}{s^{2}}\left(
\frac{2m}{\hbar ^{2}\alpha ^{2}}s^{2}-\frac{2m}{\hbar ^{2}\alpha ^{2}}\frac{%
V_{2}}{\sqrt{V_{1}}}s+4\epsilon ^{2}\right) \psi =0.
\end{equation}
Substituting $~\sigma (s)$, $~\stackrel{\sim }{\sigma }(s)~$, and $~%
\stackrel{\sim }{\tau }(s)$~in Eq.(2) we can immediately obtain ~$\pi ~$%
function as
\begin{equation}
\pi =\pm \sqrt{\frac{2m}{\hbar ^{2}\alpha ^{2}}s^{2}+(k-\frac{2m}{\hbar
^{2}\alpha ^{2}}\frac{V_{2}}{\sqrt{V_{1}}})s+4\epsilon ^{2}.}
\end{equation}
According to the NU method, the expressionin the square root must be square
of polynomial. So, one can find new possible functions for each k as
\begin{equation}
\pi =\left\{
\begin{array}{ll}
\pm (\gamma s+2\epsilon ) & ~,~for~k=\gamma ^{2}\frac{V_{2}}{\sqrt{V_{1}}}%
+4\epsilon \gamma \\
\pm (\gamma s-2\epsilon ) & ~,~for~k=\gamma ^{2}\frac{V_{2}}{\sqrt{V_{1}}}%
-4\epsilon \gamma
\end{array}
\right. ,
\end{equation}
where $~\epsilon ^{2}=-\frac{mE}{2\hbar ^{2}\alpha ^{2}}$~and $~\gamma ^{2}=%
\frac{2m}{\hbar ^{2}\alpha ^{2}}$. After determining k and ~$\pi ~$, we can
write $~\tau ~$as,
\begin{equation}
\tau =1-2\gamma s+4\epsilon .
\end{equation}
Therefore from Eqs. 4 and 14, we get exact energy eigenvalues in atomic
units as,
\begin{equation}
E_{n}=-\frac{\alpha ^{2}}{4}\left( 2n+1-\frac{V_{2}}{\alpha \sqrt{V_{1}}}%
\right) ^{2}.
\end{equation}
Using $\sigma (s)$, and $~\pi (s)$ in Eqs.$(7$ - 9$)$, one can find the
corresponding wave functions y(s) and $\phi (s),$then from Eq.(6) as
\begin{equation}
\Psi _{n}(s)=C_{n}s^{2\epsilon }e^{-\gamma s}L_{n}^{4\epsilon }(2\gamma s).
\end{equation}
Where ~$L_{n}^{\mu }(x)~$ stands for the associated Laguerre functions. One
can easily see well behavior of the wave function at infinity. As an
example, the ground state wave function behaves like
\begin{equation}
\Psi _{0}(s\rightarrow 0)\sim \left( e^{-\alpha x}\right) ^{iE_{0}/\alpha
}e^{-e^{-\alpha x}/\alpha }.
\end{equation}

Let us now consider diffrent types of the generalized Morse potential.

\subsection{\protect\bigskip\ Non-$PT$ symmetric and non-Hermitian Morse case%
}

We have considered a more general form of the complexified Morse potential
than that was previously studied~\cite{bag,ZA,S,BA}. For a special value of
its parameters, it reduces to the usual one. Thus, we may choose the
parameters from two types.

Nov, let us take the potential parameters as in the Refs.\cite{ZA} and \cite
{BQ}, $V_{1}=(A+iB)^{2}$~,$V_{2}=(2C+1)(A+iB)$,~and~$\alpha =1$, then the
potential is
\begin{equation}
V(x)=(A+iB)^{2}e^{-2x}-(2C+1)(A+iB)e^{-x},
\end{equation}
where A, B, and C are arbitrary real parameters and~$i=\sqrt{-1}$~. Such
potentials are non-$PT$- symmetric and also non-Hermitian but have real
spectra. According to Solombrino\cite{S}, this type of the complex Morse
potential is a pseudo-Hermitian and the corresponding Hamiltonian verifies
the pseudo-Hermitian propositions weakly (Proposition 3. and 5.). More
recently, the basic properties of pseudo-Hermitian operators,
pseudo-supersymmetric quantum mechanics and diagonalizable pair of
isospectral Hamiltonians with identical degeneracy structure are intensively
studied\cite{AM}.

Let us consider the real spectrum for this case. Substituting the parameters
into the energy expression, We simply get
\begin{equation}
E_{n}=-\left( n-C\right) ^{2}.
\end{equation}
In this case the spectrum is completely real and independent from the
potential parameters A and B. However, there are degeneracy for A and B. If ~%
$V_{1}~$ is real, and ~$V_{2}=A+iB~$, and ~$\alpha =i\alpha $, $PT-$
violation case, the Morse potential can be written in the following form
\begin{equation}
V(x)=V_{1}e^{-2i\alpha x}-(A+iB)e^{-i\alpha x}.
\end{equation}
For $V_{1}>0$ case, there are real spectra if and only if ~$%
\mathop{\mbox{Re}}{(V_2)}=0$. When $V_{1}<0$ there are real spectra if and
only if ~$\mathop{\mbox{Im}}{(V_2)}=0.$ It can be shown that this is related
by a pseudo-Hermitian transformation \cite{ZA,AM}.

\subsection{$PT$ symmetric and non-Hermitian Morse case}

When $~\alpha =i\alpha $, and $~V_{1}$~and ~$V_{2}~$~are real, in this case
potential takes
\begin{equation}
V(x)=V_{1}e^{-2i\alpha x}-V_{2}e^{-i\alpha x},
\end{equation}
with the $\mathop{\mbox{Re}}(V(x))=V_{1}cos(2\alpha x)-V_{2}cos(\alpha x)$
and $~\mathop{\mbox{Im}}(V(x))=-V_{1}six(2\alpha x)+V_{2}sin(\alpha x).$
From the Eq. (15) for$~V_{1}>0$~, we get no real spectra of this kind of ~$%
PT $-symmetric Morse potentials. In order to compare our results with the
ones obtained by Znojil \cite{Znojil},we simply take the parameters as $%
V_{1}=-\omega ^{2}~$, ~$V_{2}=D$, and~$\alpha =2$. In this particular case,
we get ~$\pi ~$function as
\begin{equation}
\pi =\pm \frac{i}{2}\left\{
\begin{array}{ll}
(\omega s-\alpha ) & ~,~for~k=(-iD+2\omega \alpha )/4 \\
(\omega s+\alpha ) & ~,~for~k=(-iD-2\omega \alpha )/4
\end{array}
\right.
\end{equation}
and after appropriate choice of k and ~$\pi ~$, we can write $~\tau ~$as
\begin{equation}
\tau =1-i(\omega s+\alpha ).
\end{equation}
Thus, the energy eigenvalues are reduced to the simple form
\begin{equation}
E_{n}=\left( 2n+1+\frac{D}{2\omega }\right) ^{2}.
\end{equation}
More recently many interesting properties of such particular cases were
studied by Znojil\cite{Znojil} and Bagchi and Quesne\cite{BA}.

\section{P\"{o}schl-Teller potential}

We shall consider the general form of the P\"{o}schl-Teller potential
\begin{equation}
V(x)=-4V_{0}\frac{e^{-2\alpha x}}{(1+qe^{-2\alpha x})^{2}}.
\end{equation}
This potential has more flexible form. Because it has a couple of additional
free parameters $\alpha $ and $q$ to the well known standard
P\"{o}schl-Teller form\cite{PT}$.$ First of all if we fix the free
parameters as $\alpha =1$ and $q=1$ the potential reduces to the well known
standard P\"{o}schl-Teller potential. This form of the potential was studied
extensively \ by many authors \cite{FÜ,gen}. The standard P\"{o}schl-Teller
potential was applied in the framework of the $su(2)$ vibron model\cite{CII}$%
.$ The special case of (25) for $q=1$, the modified P\"{o}schl-Teller
potential
\begin{equation}
V(x)=-\frac{D_{0}}{\cosh ^{2}(\alpha x)},
\end{equation}
is used to derive the well known $su(2)$ spectrum-generating algebra of an
infinite square well problem\cite{DL}.

Let us solve the Schr\"{o}dinger equation\ for the generalized
P\"{o}schl-Teller potential. One can get the following form with the new
variable~$s=-e^{-2\alpha x}~$ as
\begin{equation}
\psi ^{\prime \prime }(s)+\frac{1-qs}{s(1-qs)}\psi ^{\prime }+\frac{1}{%
[s(1-qs)]^{2}}[-\epsilon ^{2}q^{2}s^{2}+(2\epsilon ^{2}q-\beta
^{2})s-\epsilon ^{2}]\psi =0
\end{equation}
where $~\epsilon ^{2}=-\frac{mE}{2\hbar ^{2}\alpha ^{2}}$, and$~\beta ^{2}=%
\frac{2mV_{0}}{\hbar ^{2}\alpha ^{2}}$. Thus, one can easily get the energy
eigenvalues in atomic units \cite{he}as,
\begin{equation}
E_{n}(q,\alpha )=-\frac{\alpha ^{2}}{4}\left[ -(2n+1)+\sqrt{1+\frac{4V_{0}}{%
q\alpha ^{2}}}\right] ^{2}.
\end{equation}
The corresponding wave function becomes
\begin{equation}
\psi _{n}(s)=s^{-\epsilon }(1-s)^{\nu /2}P_{n}^{(2\epsilon ,\nu -1)}(1-2qs).
\end{equation}
Where $~\nu =1-\sqrt{1+\frac{8mV_{0}}{\hbar ^{2}q\alpha ^{2}}}$~and ~$%
P_{n}^{\mu ,\nu }(x)~$ stands for Jacobi polynomials. One can easily get
proper behavior of wave function at infinity.

\subsection{ Non-$PT$ symmetric and non-Hermitian P\"{o}schl-Teller cases}

Now, we are going to choose $V_{0}$ and $q$ as complex parameters ~ $%
V_{0}=V_{oR}+iV_{oI}~$ and$~q=q_{R}+iq_{I}$. Where $V_{oR}$, $V_{oI}$, $%
q_{R} $, $q_{I}$ and $\alpha $ arbitrary real parameters. In this case,
although the potential is complex, and corresponding Hamiltonian is
non-Hermitian and also non- $PT$-symmetric, there may be a real spectra if
and only if $~V_{0I}q_{R}=V_{0R}q_{I}$. When both parameters $V_{0},$ and $q$
are taken pure imaginary the potential turns out to be
\begin{equation}
V(x)=-4V_{0}\frac{2qe^{-4\alpha x}+i(1-q^{2}e^{-4\alpha x})}{%
(1+q^{2}e^{-4\alpha x})^{2}}.
\end{equation}
Here, we have simply used $V_{0},$ and $q$ instead of $V_{0I}$ and $q_{I}.$
The energy eigenvalues are the same given in Eqn.(28).

If q is arbitrary real parameter and $V_{0}\Rightarrow iV_{0}$ also $\alpha
\Rightarrow i\alpha $ completely imaginary, the potential becomes
\begin{equation}
V(x)=-4V_{0}\frac{(1-q^{2})\sin (2\alpha x)+i(2q+(1+q^{2})\cos (2\alpha x)}{%
(1+q^{2})^{2}+4q\cos (2\alpha x)(1+q\cos (2\alpha x)+q^{2})},
\end{equation}
and the corresponding energy eigenvalues become
\begin{equation}
E_{n}=\frac{\alpha ^{2}}{4}\left[ csgn\left( \left(
4V_{0}+iq\alpha ^{2}\right) q\alpha ^{2}\right) \left(
-\frac{B}{2}+\left( 2n+1\right) \right)
iA\\
-B(2n+1)+2+4n+4n^{2}\right] .
\end{equation}
where, we use the abbreviations as
$A=\sqrt{2\sqrt{1+(\frac{4V_{0}}{q\alpha
^{2}})^{2}}-2}$,$B=\sqrt{2\sqrt{1+(\frac{4V_{0}}{q\alpha
^{2}})^{2}}+2}.$ csgn is used for complex signum fuction in MAPLE\
Program. For a real spectrum we
take $-\frac{B}{2}+\left( 2n+1\right) =0.$ So this requires that $%
V_{0}/(q\alpha ^{2})=\pm \sqrt{n(n+1)(1+2n)^{2}}$ which is a constraint on
the potential parameters.

When all three potential parameters are complex, Hamiltonian is
non-Hermitian and also non-$~PT$-symmetric having real spectra. For
simplicity, let us take all three parameters are pure imaginary. That is $%
\alpha $ replaced by $i\alpha $, $q$ replaced by $iq$, and, $V_{0}$ replaced
by $iV_{0}.$ In this case the potential takes the form
\begin{equation}
V(x)=-4V_{0}\frac{(1+q^{2})\sin (2\alpha x)+2q+i(1-q^{2})\cos (2\alpha x)}{%
(1+q^{2})^{2}+4q^{2}(1-\cos ^{2}(2\alpha x))+4q(1+q^{2})\sin (2\alpha x)},
\end{equation}

and Hamiltonian is non-Hermitian and also non-$~PT$-symmetric. The energy
eigenvalues become
\begin{equation}
E_{n}=\frac{\alpha ^{2}}{4}\left[ \sqrt{sgn(A)}\sqrt{A}(1+2n)(-1+i)-\sqrt{|A|%
}(1+2n)(1+i)\\
-\frac{4V_{0}}{q\alpha ^{2}}+2(1+2n+2n^{2})\right] .
\end{equation}
Here there are a real spectra if and only if $A\geq 0$. sgn is used for
signum function in MAPLE Program.

\subsection{$PT$ symmetric and non-Hermitian P\"{o}schl-Teller case}

For $~PT$-symmetric and non-Hermitian potential case, we choose the
parameters ~$V_{0}$, and $q$ are arbitrarily real, and ~$\alpha \Rightarrow
i\alpha $. In this case, the potential becomes
\begin{equation}
V(x)=-4V_{0}\frac{(1+q^{2})\cos (2\alpha x)+2q+i(q^{2}-1)\sin (2\alpha x)}{%
(1+q^{2})^{2}+4q\cos (2\alpha x)(1+q\cos (2\alpha x)+q^{2})},
\end{equation}
and the energy eigenvalue is
\begin{equation}
E_{n}=-\frac{\alpha ^{2}}{4}\left[ \sqrt{|A|}(1+2n)(1+i)+\sqrt{sgn(A)}\sqrt{A%
}(1+2n)(-1+i)\\
 +\frac{4V_{0}}{q\alpha ^{2}}-2(1+2n+2n^{2})\right]
,
\end{equation}
where $A=1-4V_{0}/(q\alpha ^{2}),$ there are a real spectra if and only if $%
A=0,$ $i.e.,4V_{0}=q\alpha ^{2}$.

\section{Conclusions}

We have extended the PT-symmetric formulation, developed recently within the
non-relativistic quantum mechanics, to the more general complex Morse and P%
\"{o}schl-Teller potentials. We solved the Schr\"{o}dinger equation in one
dimension first time for the complex potentials by using Nikiforov-Uvarov
method. We studied so many different complex forms of these potentials.
Interesting features of quantum expectation theory for $PT$-violating
potentials may be affected by changing from complex to real systems. We
observed that there were some restrictions on the potential parameters for
bound states in ~$PT$~- symmetric or, more generally, in non-Hermitian
quantum mechanics. Because of the restriction $V_{0}/(q\alpha ^{2})=\pm
\sqrt{n(n+1)(1+2n)^{2}},$ there is no ground state of generalized P\"{o}%
schl-Teller potential when the parameters $V_{0}$ and $\alpha $ are pure
imaginary. Although the number of positive bound sates decreases with
increasing ~$\alpha ~$ and $q$ or decreasing $~V_{0}~$ for real family of
the P\"{o}schl-Teller potential, there are positive and negative bound
states for $~PT$~- symmetric cases. we have pointed out that our exact
resuts of complexified Morse and the P\"{o}schl-Teller potentials may
increase the applications in the study of different quantum systems


\end{document}